\newcommand{\nc}{\newcommand}
\nc{\del}{\partial}
\nc{\cF}{{\cal F}}
\nc{\cL}{{\cal L}}
\nc{\cO}{{\cal O}}
\nc{\ie}{{\it i.e}}

\nc{\rnc}{\renewcommand}
\nc{\CY}{Calabi-Yau}
\nc{\CYM}{Calabi-Yau manifold}
\nc{\CYMs}{Calabi-Yau manifolds}
\nc{\DB}{D-Brane}
\nc{\DBs}{D-Branes}
\nc{\SUSY}{supersymmetry}
\nc{\Kah}{K\"ahler}
\nc{\cs}{complex structure}
\nc{\beq}{\begin{equation}}
\nc{\eeq}{\end{equation}}
\nc{\ntwo}{${\cal N}=2$}
\nc{\nOne}{${\cal N}=1$}
\nc{\nOneStar}{${\cal N}=1^*$}
\nc{\hs}{\hspace{0.2in}}
\nc{\Z}{{\mathbb Z}}
\rnc{\P}{{\mathbb P}}
\nc{\R}{{\mathbb R}}
\nc{\C}{{\mathbb C}}
\nc{\WP}{\mathbb{WP}}
\nc{\slag}{special Lagrangian}
\nc{\cn}{\C^n}
\nc{\rn}{\R^n}
\nc{\M}{{\cal M}}
\nc{\W}{{\cal W}}
\nc{\linefour}{{\cal O}_{\P^4}(-5)}
\nc{\linen}{{\cal O}_{\P^{n-1}}(-n)}
\nc{\SO}{\hbox{SO}}
\nc{\Sp}{\hbox{Sp}}
\nc{\SU}{\hbox{SU}}
\nc{\U}{\hbox{U}}
\nc{\Tr}{\hbox{Tr}}
\nc{\Log}{\hbox{Log}}
\nc{\Cos}{\hbox{Cos}}
\nc{\Sin}{\hbox{Sin}}
\nc{\WLax}{W_{\hbox{lax}}}
\nc{\Wtree}{W_{\hbox{tree}}}

\nc{\comb}[2]{\left(\begin{array}{c}#1\\#2\end{array}\right)}
\nc{\tLambda}{\tilde{\Lambda}}

\documentclass[12pt]{article}
\usepackage{epsfig,amsmath,amsfonts,amssymb}
\begin{document}
\rightline{\vbox{\baselineskip12pt\hbox{hep-th/0312077}\hbox{7$^{\rm th}$ December, 2003}}}
\vskip 1cm
\begin{center}
  {\bf \Large \bf Effective Superpotentials, Geometry and Integrable Systems} \vskip 1cm
  {\large Kristian D. Kennaway} \\
    \vskip .2cm
    and\\
      \vskip .2cm
  {\large Nicholas P. Warner} \\
  \vskip .4cm
  {\it Department of Physics and Astronomy}\\
  {\it University of Southern California}\\
  {\it Los Angeles, CA 90089-0484, USA}
\end{center}

\begin{abstract}
We consider the effective superpotentials of \nOne\ $\SU(N_c)$ and $\U(N_c)$ supersymmetric gauge theories that are obtained from the \ntwo\ theory by adding a tree-level superpotential.  We show that several of the techniques for computing the effective superpotential are implicitly regularized by $2N_c$ massive fundamental quarks, \ie~the theory is embedded in the finite theory with nontrivial UV fixed point.  In order to study \nOne\ and \ntwo\ theories with fundamentals, we explicitly factorize the Seiberg-Witten curve for $N_f \ne 0$ and compare to the known form of the \nOne\ superpotential.  \ntwo\ gauge theories have an underlying integrable structure, and we obtain results on a new Lax matrix for $N_f = N_c$.
\end{abstract}

\section{Introduction}

Over the past few years there has been significant progress in the understanding of effective superpotentials for gauge theories with \nOne\ \SUSY, particularly \nOne\ theories that can be obtained from an \ntwo\ theory by the addition of a gauge-invariant superpotential for some of the \nOne\ superfields so as to break supersymmetry.

There are now at least four known techniques for computing the effective superpotentials in this class of theory.  The recent work goes back to \cite{Cachazo:2001jy, Cachazo:2002pr} who used string theory to motivate the computation of the superpotential in terms of period integrals of the factorized Seiberg-Witten curve.  By restricting to topological string theory, it was demonstrated by Dijkgraaf and Vafa \cite{Dijkgraaf:2002fc,Dijkgraaf:2002vw,Dijkgraaf:2002dh} that the evaluation of the superpotential reduces to a 0-dimensional matrix integral.  The reduction to matrix models was understood in field theory as a consequence of the generalized Konishi anomalies \cite{Cachazo:2002ry} and the perturbative cancellation of non-zero-momentum Feynman diagram contributions to the effective superpotential \cite{Dijkgraaf:2002xd, Aganagic:2003xq}.  In certain cases the superpotential can also be calculated using the known connection of \ntwo\ Yang-Mills theories with integrable systems \cite{Dorey:2002ad, Boels:2003fh,Hollowood:2003ds}.

In this paper we will explore the links between field theory, the geometry of Riemann surfaces, and integrable systems, in computing effective superpotentials for \nOne\ Yang-Mills theories with adjoint and fundamental matter.

We begin by revisiting the theory with a single adjoint chiral superfield $\Phi$ and no fundamental matter, which can be obtained by deforming the \ntwo\ pure gauge theory via the addition of a tree-level superpotential $W(\Phi)$.  The technique of \cite{Cachazo:2001jy} involves computing period integrals of (a reduction of) the factorized Seiberg-Witten curve of the \ntwo\ theory; this period integral is log-divergent and can be regularized by imposing a cut-off.  In previous discussions the effects of the cut-off have been ignored, because they are suppressed in the limit where it is taken to infinity.  We evaluate the effects of the cut-off when it is kept finite, and find that the corrections to the effective superpotential have the physical interpretation of  adding $N_f=2N_c$ massive chiral superfields in the fundamental representation, which regulate the UV divergences of the computational procedure by embedding the gauge theory in a UV-finite theory.  By giving a vev to these quarks, the low-energy gauge theory contains Higgs vacua in addition to the confining vacua of the $N_f=0$ theory we begin with.  Although the quarks are taken to be very massive when regulating the $N_f=0$ theory, by holomorphy the function $W$ is exact for arbitrary (non-zero) bare masses.

By evaluating the period integrals of the factorized $N_f=0$ curve, we derive a general combinatorial formula for the effective superpotential of the theory with $N_f=2 N_c$ and arbitrary $W(\Phi)$, which generalizes the known $N_f=0$ result.  By comparing to the matrix model for $N_f \neq 0$, we verify that this formula amounts to enumerating the planar diagrams of the theory with 0 and 1 boundary.  It is interesting to see the theory with $N_f=2N_c$ emerge from the $N_f=0$ calculation; as mentioned above, this means that the $N_f=0$ period integral and string theory/matrix model calculations are implicitly regularized by a model with $N_f=2N_c$, in the limit of large quark mass, which has a nontrivial conformal fixed point in the deep UV.

We are thus lead to study the \ntwo\ theory with massive fundamental hypermultiplets.  We consider the Seiberg-Witten curves of these theories and explicitly perform the factorization to the locus where all monopoles are massless.  The expression for the factorized curve generalizes the $N_f=0$ result and is also written in terms of Chebyshev polynomials.  Restricting to $N_f=N_c$, the form of the factorized curve simplifies, and we evaluate the moduli of the factorized curve for general $N$.

After breaking to \nOne\ by the addition of a tree-level $W(\Phi)$, the massless monopoles condense, and the effective superpotential is given by $W = \sum g_p \langle u_p \rangle$ where $u_p$ are the moduli of the factorized Seiberg-Witten curve.  We therefore obtain an explicit expression for the effective superpotential of the theory with $0 \le N_f < 2N_c$ massive quarks, and for $N_f=N_c$ we prove its equivalence to the expression obtained from the period integrals.

In section \ref{sec:intsys} we consider the relationship between \nOne\ superpotentials and integrable systems.  For $N_f=0$, the combinatorial formulae for $\langle u_k \rangle$ may be summarized by taking traces of powers of a single matrix, namely the scalar component of the adjoint field $\Phi$, evaluated in the vacuum of interest, which is identified with the Lax matrix of the periodic Toda chain.  In a sense, since the Lax pair completely characterizes the integrable system, this demonstrates how the integrable system emerges from the gauge theory. We obtain a simple proof of the equivalence of the algorithm of \cite{Boels:2003fh} with the period integral and factorization calculations, for $N_f=0$.

The integrable system associated to \ntwo\ SQCD was uncovered in \cite{Gorsky:1996hs,Gorsky:1996qp}, and is a particular spin chain system.  However the known Lax pair of this system is written in transfer matrix form as a chain of $2\times2$ matrices, which does not appear to be easy to work with to compute \nOne\ superpotentials.  Therefore, it would be useful to find another Lax pair for this system that takes the form of a single matrix, similar to the $N_f=0$ case.  In section \ref{sec:laxnf} we find the $N_c \times N_c$ matrix $\langle \Phi \rangle$ that encodes the $\langle u_k \rangle$ in the maximally-confining vacua, which is identified with a particular equilibrium value of the Lax matrix for the associated spin chain.

\section{Effective superpotentials from geometry}

\subsection{Computing the superpotential}
\label{sec:n1curve}

Suppose that we have pure \ntwo\ Yang-Mills theory broken to \nOne\ via a tree-level superpotential of the form:
\begin{equation}
\Wtree  ~\equiv~ \sum_{p=1}^{n+1} \,  \frac{g_p}{p}  \, {\rm Tr}  \big(\Phi^p\big)
~\equiv~ \sum_{p=1}^{n+1} \,    g_p   \, u_p \,.
\end{equation}
In  \cite{Cachazo:2001jy}, string theory arguments were given for the computation of the glueball superpotential in terms of periods of the differential form (``resolvent''): 
\begin{eqnarray}
\omega(x) &=& \frac{1}{2} \left(W'(x) - \sqrt{(W'(x))^2 ~+~ f_{n-1}(x) } \right)\, dx \nonumber \\
&\equiv& \frac{1}{2}(W'(x) - y(x)) dx
\end{eqnarray}
which is single-valued on the genus $n-1$ Riemann surface $y^2 = W'(x)^2 + f_{n-1}(x)$ \footnote{This curve may also be obtained by factorizing the Seiberg-Witten curve of the associated \ntwo\ theory obtained when $W=0$, and discarding the roots of the curve corresponding to condensed monopoles; we will use this technique in section \ref{sec:sw}.}.  The compact $A$-periods yield the gaugino condensates, $S_i$, while the non-compact $B$-periods, $\Pi_i$ yield the derivatives of the free energy $\frac{\del \cF}{\del S_i}$.  We choose the branches of the square root so that on the first sheet $\omega(x)$ vanishes in the classical limit $f_{n-1} \rightarrow 0$; therefore on the second sheet $\omega(x) \rightarrow W'(x)$.

In this paper we will focus on the maximally-confining phase of the theory, for which the resolvent degenerates:
\begin{equation}
\label{confW}
y(x)  ~=~ \sqrt{(W'(x))^2 ~+~ f_{n-1}(x) } \, dx ~=~ G_{n-1}(x)  \, \sqrt{(x-c)^2 ~-~
\mu^2} \, dx\,, 
\end{equation}
for some polynomial, $G_{n-1}(x)$ of degree $(n-1)$.   For $\U(N)$ theories, it is convenient to use the freedom to shift $x$ so as to set $c=0$; this is not allowed for $\SU(N)$, for which the center of the cut is not a free parameter, but the $\SU(N)$ results may be obtained from the $\U(N)$ at the end of the calculation by decoupling the overall $\U(1)$ trace (we will come back to this point later).  The gaugino condensate is then given by:
\begin{equation}
\label{Scomp}
S ~=~\frac{1}{2 \pi \imath} \oint_A \omega(x) =  \pm \frac{1}{4 \pi \imath} \oint_A y(x) ~=~\pm \frac{1}{2 \pi \imath} \int_{-\mu}^{\mu} \,
G_{n-1}(x)  \, \sqrt{x^2 ~-~ \mu^2} \, dx 
\end{equation}
where the sign depends on the orientation of the contour.  The $B$-period is given by integrating along a contour from infinity on the second sheet, through the cut to infinity on the first sheet.  The logarithmic divergence of this integral needs to be regularized, and this is usually done by a introducing a UV cut-off: 
\begin{equation}
\label{Picomp}
\Pi_B  ~=~ \int_B \omega ~=~ \int_{x_- = \Lambda_0}^{x_+  = \Lambda_0}  \omega   
= - \int_\mu^{\Lambda_0} G_{n-1}(x)  \, \sqrt{x^2 ~-~ \mu^2} \, dx  \,,
\end{equation}
where $x_-$ and $x_+$ denote the values of $x$ on the lower and upper sheets respectively.  
The effective superpotential is then given by:
\begin{equation}
\label{eq:weff}
W_{\rm eff}   ~=~ N \, \Pi_B + N W(\Lambda_0) ~+~ \alpha \, S
\end{equation}
where $\alpha =  2 \pi \imath \tau_0$ is the bare gauge coupling, and the second term is added to cancel the contribution from the upper limit of the integral in $\Pi_B$.  The effect of the $\alpha$ term is to combine with the log-divergent piece of $\Pi_B$ to give the (finite) dynamical scale of the theory \cite{Cachazo:2001jy}.

In computing the effective superpotential by this method, the approach taken in the recent literature is to send $\Lambda_0 \rightarrow \infty$, causing its effects to decouple from the theory.  However, it is interesting to study the effects of keeping the cut-off finite.  We will henceforth take the cut-off to be large but finite, and investigate the effects on the low-energy gauge theory; this amounts to keeping the $O(1/\Lambda_0)$ terms in $\Pi_B$ and subsequent calculations.

\subsection{Example: U(2)}
\label{sec:periodu2}

Before analysing $U(N)$, for general $N$, we consider the simplest example of $U(2)$ with a tree-level mass: $W = \frac{1}{2} m \Tr \Phi^2$.  The effective 1-form is

\begin{equation}
y(x) = m \sqrt{x^2-\mu^2}
\end{equation}
which is single-valued on a two-sheeted Riemann surface with a cut
between $x = \pm \mu$.  The glueball is given by the A-period:

\begin{equation}
 S = \frac{1}{4 \pi \imath} \oint_A y(x) dx = \frac{1}{2 \pi \imath} \int_{-\mu}^{\mu} y(x) dx =  \frac{1}{4} m \mu^2
\end{equation}
and the B semi-period is
\begin{eqnarray}
\Pi_B &=&- \int_\mu^{\Lambda_0} y(x) dx \nonumber \\
&=&- \frac{m}{2} \left( \pm \Lambda_0^2 \sqrt{1 - \frac{\mu^2}{\Lambda_0^2}} + \mu^2 \Log\left(\frac{\mu}{\Lambda_0\Big (1\pm\sqrt{1 - \frac{\mu^2}{\Lambda_0^2}}\Big)}\right)\right) \nonumber \\
&=& \mp \frac{m \Lambda_0^2}{2} \sqrt{1-\frac{4S}{m \Lambda_0^2}} - S \Log\left(\frac{S}{\frac{m \Lambda_0^2}{2}\Big(1\pm\sqrt{1-\frac{4S}{m \Lambda_0^2}}\Big)-S}\right) \nonumber \\
\end{eqnarray}
where the integral is evaluated using hyperbolic functions and the two branches come from ${\hbox{sinh}}(x) = \pm \sqrt{{\hbox{cosh}^2(x)}-1}$ (this amounts to a choice of contour, \ie~integrating to the point above $\Lambda_0$ on one of the two sheets).  As mentioned in the previous section, the role of $\alpha$ in (\ref{eq:weff}) is to replace the $N \Log(m \Lambda_0^2)$ term in $\Pi_B$ by the finite scale $N \Log(\Lambda^3)$. This may be implemented in practice by setting $\alpha = N \Log(\frac{\Lambda^3}{m \Lambda_0^2})$ in (\ref{eq:weff}).

We find

\begin{equation}
\label{eq:u2plus}
W = NS \Big(1- \Log(\frac{S}{\Lambda^3}) \Big) - \frac{S^2}{m \Lambda_0^2} - \frac{2 S^3}{2 (m \Lambda_0^2)^2} - \frac{5 S^4}{3 (m \Lambda_0^2)^3} - \frac{14 S^5}{4 (m \Lambda_0^2)^4} - \ldots
\end{equation}
Therefore in the limit $\Lambda_0 \rightarrow \infty$ (equivalently, keeping $\Lambda_0$ finite and considering energies $m << \Lambda_0$) the infinite correction series tends to zero and the glueball superpotential (\ref{eq:u2plus}) reduces to the usual Veneziano-Yankielowicz superpotential.

The form of the series (\ref{eq:u2plus}) is the same as that obtained for $U(2), N_f=4$, with $\Lambda_0$ identified with the quark mass.  The known formula for $W(S)$ with tree-level superpotential $W = \frac{1}{2} m \Tr \Phi^2 + \sum_{i=1}^{N_f} \mu \tilde{Q}_i Q^i + \tilde{Q}_i \Phi^i_j Q^i$ is \cite{Argurio:2002xv,Brandhuber:2003va}

\begin{equation}
W(S) = N_c S (1- \Log(\frac{S}{m \Lambda_0^2}) ) - N_f S \Log(\frac{\mu}{\Lambda_0}) - N_f S \left(\frac{1}{2} + \frac{\sqrt{1 - 4 \alpha S} - 1}{4 \alpha S} - \Log(\frac{1+\sqrt{1 - 4 \alpha S}}{2})\right)
\end{equation}
with $\alpha = 1/(m \mu^2)$.  Setting $N_c=2, N_f=4, \mu = \Lambda_0$ and performing  the  series 
expansion, we recover the   expression in (\ref{eq:u2plus}).

Choosing the other branch of $\Pi_B$ we obtain the negative of (\ref{eq:u2plus}).  This branch describes a Higgs branch \cite{Brandhuber:2003va}, where the gauge symmetry is broken by giving a vev to the scalar component of the quark superfields (an arbitrary Higgs vacuum can be obtained by writing $W = \alpha S +  \sum_{i=1}^{N} \Pi_B$ and choosing the branch of $\Pi_B$ termwise, \ie~for each period integral we choose whether to integrate along a contour on the first or second sheet).

We will show in Section \ref{sec:uv} that this holds true for general $N_c$ and $W(\Phi)$, and the corrections obtained by keeping the cut-off dependence in the period integral indeed have the physical interpretation of $N_f = 2 N_c$ massive quark superfields, which serve to regularize the divergences of the calculation.

If instead of $\U(N)$ gauge theory we considered $\SU(N)$, the foregoing discussion would be modified by the need to ensure ``quantum tracelessness" of the vacuum, \ie~that $\langle u_1 \rangle = 0$.  This may be achieved by taking the tree-level superpotential $W = \frac{1}{2} m \Tr \Phi^2 + \lambda \Tr \Phi$ and proceeding with the above analysis, treating $\lambda$ as a Lagrange multiplier to enforce $\langle u_1 \rangle = \langle \Tr \Phi \rangle = 0$.  This has the effect of shifting the center of the cut away from the origin.  Instead of repeating the calculation for $\SU(2)$, we will defer until later when we consider the general $\U(N)$ and $\SU(N)$ cases.

\subsection{Evaluation of the period integral for general W}
\label{sec:generalper}

The period integrals, (\ref{Scomp}) and (\ref{Picomp}), are elementary but one can obtain a simple closed form in terms of $\Wtree$.  This can be evaluated and gives a combinatorial formula for the moduli $u_k$ which can be compared to other techniques.  Make the change of variables\footnote{We again assume that $x$ has been centered on the cut; we will discuss later how to include the effects of a shift.}:
\begin{equation}
\label{newvar}
x ~=~ \frac{1}{2} \,\mu \,  ( \xi ~+~ \xi^{-1}) \,,
\end{equation}
and define series expansions:
\begin{eqnarray}
 \label{Wser}
W \big(\frac{1}{2} \,\mu \,  ( \xi ~+~ \xi^{-1}) \big) &=&  b_0 ~+~
 \sum_{k= 1}^{n+1} ~ b_k \, ( \xi^k ~+~ \xi^{-k}) \,, \\
W'\big(\frac{1}{2} \,\mu \,  ( \xi ~+~ \xi^{-1})\big)  &=& c_0 ~+~ \sum_{k=1}^n ~
c_k \, ( \xi^k ~+~ \xi^{-k})
 \label{Wprimeser}
 \end{eqnarray}
Note that the series take this form because of the symmetry of (\ref{newvar}) under $\xi \to \xi^{-1}$.  Under this change of variables the integrand may be written:
\begin{eqnarray}
\label{magicident}
\frac{1}{2} \,\mu\, ( \xi ~-~ \xi^{-1})  \,   G_{n-1}\Big(\frac{1}{2} \,\mu\, ( \xi ~+~ \xi^{-1}) \Big)  
&=& G_{n-1}(x)  \, \sqrt{x^2 ~-~ \mu^2}    \,, \\
&=& W'(x) \, \sqrt{1 ~+~ \frac{f_{n-1}(x) }{ (W'(x))^2 }}  \\ 
&=&W'(x)  ~+~\cO(\xi^{-1}) \,.
\end{eqnarray}
The left-hand side is manifestly odd under $\xi \to 1/\xi$, while the right-hand side shows that all the non-negative powers in the $\xi$-expansion are given by (\ref{Wprimeser}).  It therefore follows that under the change of variables, one has 
\begin{equation}
\label{integrandseries}
\sqrt{(W'(x))^2 ~+~  f_{n-1}(x)} ~=~  G_{n-1}(x)  \, \sqrt{(x^2 -\mu^2)}
~=~ \sum_{k=1}^n ~ c_k \, ( \xi^k ~-~ \xi^{-k})  \,. 
\end{equation}
Note in particular that the left-hand side of (\ref{magicident}) is manifestly odd under $\xi \to 1/\xi$, therefore $c_0 =0$ in (\ref{Wprimeser}).

Define  $[\dots]_-$ to mean: discard all the  non-negative powers of $\xi$ in $[\dots]$. We  may then write the last equation as:
\begin{equation}
\label{niceform}
\sqrt{(W'(x))^2 ~+~  f_{n-1}(x)} ~=~ W'\Big(\frac{1}{2} \,\mu\, (\xi + \xi^{-1})
\Big) ~-~ 2\, \Big[ W'\Big(\frac{1}{2} \,\mu\, ( \xi +  \xi^{-1}) \Big)\Big]_- \,.
\end{equation}
One can now easily perform the integrals (\ref{Scomp}) and (\ref{Picomp}).  The former is simply given by taking $\xi =e^{\imath \theta}$ for $0 \le \theta \le \pi$, and it picks out the $\xi$-residue:
\begin{equation}
\label{Sres} 
S~=~   \frac{\mu}{2}   ~ c_1 
\end{equation}
To perform the second integral we first note that:
\begin{equation}
\frac{d }{d \xi}\, \Big[ W \big(\frac{1}{2} \,\mu\, 
( \xi +  \xi^{-1}) \big) \Big]_-  ~=~   ~-~ 
\frac{1}{2} \,\mu\,  c_1\,   \xi^{-1}~+~ \frac{1}{2}  \,\mu\, (1 - \xi^{-2}) \,  
\Big[ W' \big(  \frac{1}{2} \,\mu\, (\xi +    \xi^{-1})  \big) \Big]_-  \,,
\end{equation}
and therefore:
\begin{eqnarray}
\int  \sqrt{(W'(x))^2 ~+~  f_{n-1}(x)}\, dx  & = &
- \mu\,  c_1\,  \Log(\xi) ~+~ W(x) - 2 \Big[ W(x) \Big]_-  
\label{indefinta}
\\ & = & - \mu\, 
c_1\, \Log(\xi) ~+~b_0~+~  \sum_{k=1}^n ~ b_k \, ( \xi^k ~-~ \xi^{-k}) \,,
\label{indefintb}
\end{eqnarray}
where $x = \frac{1}{2} \,\mu\, (\xi + \xi^{-1})$. To obtain $\Pi$, we must evaluate this between $\xi =1$ and $\xi = \xi_0$, where
\begin{equation}
\label{xizerodefn}
\xi_0 \equiv \xi(\Lambda_0) ~=~ \frac{\Lambda_0}{\mu} \, \bigg(1 ~+~ \sqrt{1 - \Big(\frac{\mu}{\Lambda_0}\Big)^2} \, \bigg)  \,.
\end{equation}
This yields:
\begin{equation}
\label{Pians}
\Pi ~=~    b_0 ~+~    \mu\,  c_1\,  \Log(\xi_0) ~-~  \Big( W(\Lambda_0)   - 2 \big[ W(x) \big]_- \Big|_{\xi =\xi_0} \Big) \,,
\end{equation}
where the  definite integral has been evaluated using (\ref{indefinta}) at $\xi = \Lambda_0$
and using (\ref{indefintb}) at $\xi = 1$.

In the limit of large $\Lambda_0$ the last term in (\ref{Pians}) vanishes since it only involves negative powers of $\xi_0 \sim \Lambda_0^{-1}$. Taking this limit, and using (\ref{Sres})  one obtains:
\begin{equation}
\label{eq:periodpuregauge}
\Pi ~=~   b_0 ~+~  2\, S \,\Log\Big(\frac{2 \Lambda_0}{\mu} \Big) ~-~ W(\Lambda_0) \,.
\end{equation}
Therefore
\begin{eqnarray}
\label{eq:glueballper}
W &=& N b_0 + 2 N S \Log\Big(\frac{2 \Lambda}{\mu} \Big) \nonumber \\
\end{eqnarray}
It can be shown that for general $\Wtree(\Phi)$, (\ref{eq:glueballper}) can be extremized with respect to $S$ by taking $\mu = 2 \Lambda$, and we find

\begin{equation}
\label{eq:glueballlow}
W_{\hbox{low}}= N_c \sum_{p=1}^{\lfloor \frac{n}{2} \rfloor} \frac{g_{2p}}{2p} \comb{2p}{p} \tilde \Lambda^{2p}
\end{equation}
where we have evaluated the coefficients $b_0$ in the series expansion (\ref{Wser}). 

The engineering of \nOne\ gauge theories from string theory \cite{Cachazo:2001jy} involves D-branes on generalized conifold geometries.  From the string theory perspective it is tempting to also interpret the cut-off in terms of branes.  That is, it is really only physically natural to terminate the period integral on another brane.  Having a stack of $M$ branes at $\Lambda_0$ would mean that one started with a larger (product) gauge group and that the original $\SU(N)$ theory is actually coupled to $M$ bi-fundamental matter multiplets with a (gauged) $\SU(M)$ ``flavour'' group (see \cite{Hofman:2002bi} for an analysis of this theory).  However, when the second set of branes become non-compact, their associated gauge coupling tends to zero, and the $\SU(M)$ gauge factor becomes a global $\SU(M)$ flavour symmetry.  Thus, string theory suggests that keeping the UV cut-off terms should yield the superpotential associated with the coupling to fundamental matter multiplets.  This is indeed what we find in explicit calculations.

In other words, we claim that the term
\begin{equation}
\label{mattercont}
2 N \Big[ W \Big(\frac{1}{2} \,\mu\, (\xi + \xi^{-1}) \Big) \Big]_- \Big|_{\xi =\xi_0} \,,
\end{equation}
summarizes the contribution of  these additional matter multiplets.  We will prove this in the following section.

If one also recalls that the canonical form of the $B$-period integral, (\ref{Picomp}), involves
an integral from the lower to the upper sheet of the Riemann surface, then this extra term 
may be thought of arising  from $N_c$ branes  (or anti-branes) at each limit.   Thus one
can also extract  the results for $N_f =N_c$ by regulating the upper and lower limits independently.
We will develop and extend this observation in the next section.

\section{Effective superpotentials for SQCD with one adjoint and $N_f$ fundamental quarks}

In this section we consider the \nOne\ theories with matter content of \ntwo\ SYM with massive fundamental hypermultiplets, \ie~a single adjoint $\Phi$ and $N_f$ fundamental quark and antiquark multiplets.

It was shown in \cite{Dijkgraaf:2002xd, Aganagic:2003xq} that for the class of \nOne\ gauge theories considered here, contributions to the effective superpotential come only from the zero-momentum planar diagrams of the theory, which are counted by an associated matrix integral.  In this section we use this matrix model to derive the generating function of planar diagrams for the theory with $N_f$ massive quarks, and recover  the known expression for the quark contributions to the superpotential in terms of period integrals of the spectral curve.

It is then easy to see how taking $N_f=2N_c$ fundamental quarks causes the divergences of the $N_f=0$ period integral to cancel, and therefore the massive quarks regularize the superpotential computation. 

\subsection{Effective superpotentials from matrix models}

The glueball superpotential $W(S)$ has been studied using matrix models in a large number of papers.  In particular, the $\U(N)$ theory with tree-level superpotential

\begin{equation}
\label{su2flavor}
\Wtree = W_\Phi(\Phi) + \mu \tilde Q Q + g \tilde Q \Phi Q
\end{equation}
where $Q$ is a fundamental chiral superfield and $\tilde Q$ its conjugate, has been studied for $W_\Phi(\Phi) = \frac{1}{2} m \Tr \Phi^2$ in \cite{Argurio:2002xv, Brandhuber:2003va}.  We will derive the solution to the matrix model for a general $W_\Phi$ using the combinatorics of planar diagrams, focusing on the contributions of the quarks to the effective superpotential.

It can be shown that perturbative contributions to the large-$N$ matrix model come from planar diagrams with 0 and 1 quark boundary:

\begin{equation}
W(S) = N_c \frac{\partial {\cal F}_{\chi=2}}{\partial S} + N_f {\cal F}_{\chi=1}
\end{equation}
Contributions to the first term come only from $\Phi$ self-interactions, so their combinatorics are the same as for the theory without quarks.  Diagrams with one external boundary can be counted by decomposing the counting problem into two parts: the combinatorics of the $\Phi$ diagrams on the interior of the disc, and the combinatorics of the boundary of the disc.

The first problem is equivalent to counting the planar $n$-point Green's functions $G_n(g_i)$ of the theory without quarks (\ie~planar $\Phi$ diagrams -- possibly disconnected -- with $n$ external $\Phi$ legs).  This problem was solved in \cite{Brezin:1978sv}, as follows:

By definition,

\begin{equation}
G_n(g_i)=\langle \Tr \Phi^n \rangle = \int_{a}^{b} d\lambda\,  y(\lambda)\,  \lambda^n
\end{equation}
where the second equality follows from the change of variables from the matrix integral to the eigenvalue basis and $a, b$ are the endpoints of the eigenvalue branch cut.  In other words, the sum of the planar Greens functions at each order are given by the corresponding moment of the eigenvalue density $y(\lambda)$.

The generating function for the Greens functions is
\begin{equation}
\label{genfn}
\phi(j) = \sum_{k=0}^{\infty} j^k G_k = \frac{1}{j} \omega(\frac{1}{j})
\end{equation}
where the second equality is given in terms of the resolvent
\begin{equation}
\omega(\lambda) = \frac{1}{2}(W'(\lambda) - \sqrt{W'(\lambda)^2 +
  f_{n-1}(\lambda)})
\end{equation}
by summing the geometric series and converting to a contour integral, using the fact that the eigenvalue density $\rho(\lambda)$ is equal to the discontinuity in $\frac{1}{2 \pi \imath} \omega(\lambda)$ across the branch cut, and has asymptotic behaviour $\omega(x) \sim 1/x$ as $x \rightarrow \infty$.

To include the combinatorics of the boundary requires multiplying by $\frac{(k-1)!}{k!} = \frac{1}{k}$ at order $k$ in the expansion of $G$, to take into account the $(k-1)!$ distinct ways to connect a boundary quark with a leg of the internal Greens function\footnote{At first sight, it looks like an arbitrary connection of a boundary leg to an internal leg can make the overall graph non-planar, however we can always perform a corresponding crossing operation on the internal part of the diagram to undo this non-planarity.}, and the $\frac{1}{k!}$ coming from the expansion of $e^S$ to order $k$.  The factor $\frac{1}{k}$  can be incorporated into  (\ref{genfn}) simply by integrating it:
\begin{eqnarray}
\Pi(j) &=& \int{\frac{1}{j^2} \omega(\frac{1}{j})} dj \nonumber \\
&=& - \int{\omega(x) dx} \nonumber \\
&=& -\frac{1}{2} \int(W'(x) - \sqrt{W'(x)^2 + f_{n-1}(x)} ) dx
\end{eqnarray}
where we have changed variables $x = \frac{1}{j}$ and used the definition of $\omega(x)$.  

The factors of $j$ count the number of external legs of the Greens function; therefore terms of order $j^k$ are associated to $k$ powers of the Yukawa coupling $g$, and $k$ quark propagators $\frac{1}{M}$ to connect up the $k$ external quarks on the boundary.  Therefore the one-boundary contribution to the matrix integral is given by

\begin{equation}
\label{eq:quarkgenfn}
-\frac{1}{2} \int_M^{\Lambda_0}(W'(x) - \sqrt{W'(x)^2 + f_{n-1}(x)}) dx
\end{equation}
where we have regularized the log-divergent integral by a cut-off $\Lambda_0$ and set $g=1$ by rescaling $Q$ and $\tilde Q$.  This is the usual formula for the contribution of a fundamental field of mass $M$ to the effective superpotential \cite{Cachazo:2001jy,Naculich:2002hr,Cachazo:2003yc}.

\subsection{UV cut-off as regularization by $N_f=2N_c$ fundamental quarks}
\label{sec:uv}

As mentioned in section \ref{sec:n1curve}, the effective superpotential for the $N_f=0$ theory (in a maximally confining vacuum) is given by

\begin{equation}
W \sim 2 N_c  \int_{\mu}^{\infty} \omega + \alpha S
\end{equation}
where the integral is formally divergent and is usually cut off at a point $\Lambda_0$.  Introducing $N_f$ fundamentals gives the (again formally divergent) contribution

\begin{equation}
\label{eq:wnf}
W_{N_f} \sim - \sum_{i=1}^{N_f} \int_{m_i}^{\infty} \omega
\end{equation}

However, when $N_f=2N_c$, the contours combine and the integration domains are now finite, so the divergence of the integrals have been regularized.  When all $m_i$ are equal we may write $m_i \equiv \Lambda_0$ and we can explicitly see the role of the $2N_c$ fundamental fields in implementing the cut-off of the $N_f=0$ integral: they act as regulators for the UV divergences of the calculation, by embedding the gauge theory in a UV-finite theory.

Note that the same divergent integral appears in the evaluation of the $N_f=0$ matrix model.  Therefore, the matrix model is also implicitly regularized by $2N_c$ quarks.  It has been argued \cite{Aganagic:2003xq} that the matrix model (and string theory) calculation is naturally embedded in $\U(N+k|k)$ (this is the origin of the discrepancies that can appear in certain cases in the matrix model compared to the gauge theory).  This is an additional UV completion that is not necessary from the field theory approach of taking \ntwo\ Yang-Mills for a specific gauge group ${\cal G} = \U(N)$, factorizing the curve and deforming to \nOne.  Since the matrix model calculation still involves cutting off a log-divergent period integral, this is also implemented physically by introducing $N_f=2N_c$ massive quarks.  In other words, there are two UV completions needed to make the matrix model well-defined in the UV.

As we have seen, when $\Lambda_0$ is taken to be large but finite, it gives finite (but small) corrections to the expression for the glueball superpotential $W(S)$.  Therefore, the vacuum expectation value for the glueballs $\langle S_i \rangle$ will be perturbed from that of the theory we started with (\nOne\ Yang-Mills theory with a massive adjoint and no fundamental matter).  In other words, in terms of the \nOne\ curve, the presence of the cut-off at a finite distance from the cuts cause the size and center of the cuts to be perturbed.  Because of this deformation, this \nOne\ curve {\it cannot} be obtained by factorizing the SW curve of pure \ntwo\ Yang Mills.

Therefore, in regularizing the $N_f=0$ theory by imposing a finite cut-off on the divergent integral, we have gone off-shell (\ie~the vacua of this theory do not solve the equations of motion of the $N_f=0$ theory).  Physically, this amounts to embedding the $N_f=0$ theory in a larger theory with $N_f=2N_c$ massive quark flavours.  It is only in the limit of infinite quark mass (infinite cut-off) that the effects of the quarks on the vacuum structure of the theory decouple and we approach the on-shell vacua of the $N_f=0$ theory.

In practice we can think of the effective superpotential for $0\le N_f \le 2N_c$ fundamentals (as computed using the matrix model or using the technique of \cite{Cachazo:2001jy}), as always being generated by the UV-finite theory with $2N_c$ fundamental fields, with masses that are either kept finite or which are taken to infinity at the end of the calculation and decouple from the theory.  In other words, if we have $\tilde N_f$ fundamental fields of finite mass, then the remaining $2N_c-\tilde N_f$ are of mass $\Lambda_0 \gg m$.  Therefore:

\begin{eqnarray}
W &\sim& 2 N_c \int_{\mu}^{\infty} \omega -  \sum_{i=1}^{\tilde N_f} \int_{m_i}^{\infty} \omega - (2N_c-\tilde N_f) \int_{\Lambda_0}^{\infty} \omega \nonumber \\
&=& 2 N_c \left(  \int_{\mu}^{\infty} \omega - \int_{\Lambda_0}^{\infty} \omega \right) + \sum_{i=1}^{\tilde N_f}\left( \int_{\Lambda_0}^{\infty} \omega -  \int_{m_i}^{\infty} \omega \right) \nonumber \\
&=&  2N_c  \int_{\mu}^{\Lambda_0} \omega -    \sum_{i=1}^{\tilde N_f} \int_{m_i}^{\Lambda_0} \omega
\end{eqnarray}
and all integrals are finite.  We can then decouple the quarks of mass $\Lambda_0$ by taking $\Lambda_0 \rightarrow \infty$, and using the results of section \ref{sec:generalper} we find the expression for $W$

\begin{eqnarray}
\label{eq:generalnf}
W &=& N_c \left( b_0 + \mu c_1 \Log(\frac{2 \Lambda_0}{\mu}) \right) + \nonumber \\
&& \hspace{0.5in} \frac{1}{2}\sum_{i=1}^{N_f} (\mu c_1 \Log(\frac{ m}{\Lambda_0}) + 2 [W(\xi(m_i))]_-) + \alpha S \nonumber \\
&=& N_c b_0 + \frac{1}{2} \mu c_1 \Log(2^{2N_c} \frac{\Lambda_0^{2N_c-N_f} \prod_{i-1}^{N_f}m_i}{\mu^{2N_c}} ) + \sum_{i=1}^{N_f} [W(\xi(m_i))]_- + \alpha S \nonumber \\
&=& N_c b_0 + N_c \mu c_1 \Log(\frac{2 \tilde{\Lambda}}{\mu} ) + \sum_{i=1}^{N_f} [W(\xi(m_i))]_-
\end{eqnarray}
In other words, we have shown that the contribution to $W$ of a quark of mass $m_i$ is ${[W(\xi(m_i))]}_-$, as claimed.

We should expect to recover the result (\ref{eq:generalnf}) by factorizing the SW curve for \ntwo\ Yang-Mills with $N_f$ massive hypermultiplets.  In the following section we will solve the factorization problem for general $N_f$ and prove the equivalence of the resulting superpotential to the expression (\ref{eq:generalnf}) for  $N_f=N_c$.

\subsection{Effective superpotentials from Seiberg-Witten theory}
\label{sec:sw}

In previous sections we studied the vacua of the \nOne\ gauge theory directly.  These results descend from the structure of the underlying \ntwo\ theory one obtains by setting $W=0$, so we turn our attention now to the \ntwo\ $\U(N)$ gauge theories with $N_f$ fundamental hypermultiplets.

As is well-known, the vacuum structure of \ntwo\ gauge theories are described by a fibration of a Riemann surface (the Seiberg-Witten curve) over the moduli space.  At points in the moduli space where the curve degenerates, physical degrees of freedom (monopoles, dyons or W-bosons) become massless.  

Written in \nOne\ language, the effective superpotential for the \ntwo\ theory in the neighbourhood of a massless monopole point is \cite{Seiberg:1994aj,Seiberg:1994rs}

\begin{equation}
W(M_m, \tilde{M}_m, u_p, \Lambda) = \sum_{m=1}^{N-1} \tilde{M}_m M_m a_{D,m}(u_p, \Lambda)
\end{equation}
where $M_m$ are the monopole hypermultiplets, $a_{D,m}$ are the periods of the SW curve that determine the monopole masses, and $u_p$ are the gauge-invariant curve moduli $\frac{1}{p} \Tr \Phi^p$ that parameterise the vacua of the \ntwo\ theory.  After breaking to \nOne\ by the addition of a tree-level superpotential, the Intriligator-Leigh-Seiberg linearity principle \cite{Intriligator:1994jr} implies that the exact superpotential becomes

\begin{equation}
\label{eq:wmonopole}
W(M_m, \tilde{M}_m, u_p, \Lambda, g_p) = \sum_{m=1}^{N-1} \tilde{M}_m M_m a_{D,m}(u_p, \Lambda) + \sum g_p u_p
\end{equation}

The equation of motion for the monopole fields imposes that $a_{D,m} = 0$.  This is true iff the corresponding B-cycle of the Seiberg-Witten curve degenerates, therefore the maximally-confining vacua (where the gauge symmetry is classically unbroken) correspond to the point in the \ntwo\ moduli space where all the monopoles become massless, and the SW curve degenerates completely\footnote{The factorized curve, after discarding the double roots associated to the massless monopoles that have been condensed, is identified with the \nOne\ curve $y^2 = W'(x)^2 + f_{n-1}(x)$ studied in section \ref{sec:n1curve}.}.  The equation of motion for the $u_p$ implies that there is a nonzero monopole condensate in the confining \nOne\ vacua, \ie~confinement of the \nOne\ theory is associated to monopole condensation.

After evaluating (\ref{eq:wmonopole}) at the factorization locus, the exact effective superpotential then becomes
\begin{equation}
W_{\hbox{low}}(g_p, u_p, \Lambda) = \sum g_p u_p|_{\{a_{D,m} = 0\}}
\end{equation}
Thus, evaluation of the effective superpotential is equivalent to solving the factorization of the spectral curve.  Once we know the moduli $u_p$ at the factorization locus we can immediately read off the effective superpotential corresponding to any given $\Wtree$.

For $N_f=0$ the factorization of the curve is achieved as follows \cite{Douglas:1995nw}:
\begin{eqnarray}
\label{curvepure}
y^2 &=& P_N(x)^2 - 4 \Lambda^{2N}\nonumber \\
&=& 4 \Lambda^{2N}( T_N(x)^2 - 1)
\end{eqnarray}
where $P_N(x) = x^N + \sum_{i=1}^{N} s_i x^{N-k}$, with $s_1=0$ for the $\SU(N)$ curve, and $T_N(x)$ are the Chebyshev polynomials of the first kind, defined by

\begin{eqnarray}
\label{eq:cheby}
T_N(x \equiv \Cos(\theta))&=& \Cos(N \theta) \nonumber \\
&=& \frac{N}{2} \sum_{r=0}^{\lfloor \frac{N}{2} \rfloor} \frac{(-1)^r}{N-r} \comb{N-r}{r}(2 x)^{N-2r}
\end{eqnarray}
which gives the expansion of $\Cos(N \theta)$ in terms of $\Cos(\theta)$.  In other words, by tuning the parameters $s_k$ of the curve (equivalently, the gauge-invariant moduli $u_k = \frac{1}{k} \Tr \Phi^k$, which are related to the $s_k$ via $k u_k + k s_k + \sum_{i=1}^{k-1} i u_i s_{k-i} = 0$), we can obtain $P_N(x) =2  \Lambda^N T_N(\frac{x}{2 \Lambda})$, therefore 

\begin{equation}
P_N(x)^2 - \Lambda^{2N} = \Lambda^N(\Cos^2(N \theta) - 1) = \Lambda^N(\Sin^2(N \theta)) = \Lambda^N \sqrt{1-\frac{x^2}{4\Lambda^2}} U_{N-1}(\frac{x}{2 \Lambda})^2
\end{equation}
where $U_N(x)$ are the Chebyshev polynomials of the second kind, given by

\begin{equation}
U_N(x) = \sum_{r=0}^{\lfloor \frac{n}{2} \rfloor} (-1)^r \comb{n-r}{r} (2x)^{n-2r}
\end{equation}

From (\ref{eq:cheby}) one can read off the values of the $s_k$ in this vacuum.  To convert to $u_k$ we use  the product form

\begin{equation}
T_N(x) = 2^{N-1} \prod_{k=1}^N (x-\Cos( \frac{(2k-1) \pi}{2N}) \equiv 2^{N-1} \prod_{k=1}^N (x-x_k) 
\end{equation}
with

\begin{equation}
\label{eq:powersum}
u_k = \frac{1}{k} \sum_{i=1}^N x_i^k
\end{equation}
Expanding the power sum for $\SU(N)$ gives

\begin{equation}
u_k = \left\{ \begin{array}{ll}
0 & k \hbox{ odd} \\
\frac{1}{k} \comb{k}{k/2} \Lambda^{k} & k \hbox{ even}
\end{array} \right. 
\end{equation}
and therefore we have the effective superpotential 

\begin{eqnarray}
\label{eq:wpurelow}
W &=& \sum g_i \langle u_i \rangle \nonumber \\
&=& \sum \frac{g_{2k}}{2k} \comb{2k}{k} \Lambda^{2k}
\end{eqnarray}

The result for $\U(N)$ may be obtained from (\ref{eq:wpurelow}) by shifting $x \rightarrow x + u_1/N = x - \phi$ in (\ref{eq:powersum}) to account for the non-zero trace of $\Phi$, where the equality follows since we are in a maximally-confining $U(N)$ vacuum, for which classically $\langle \Phi \rangle = {\hbox{diag}}(\phi, \phi, \ldots, \phi)$.  Explicitly, for the maximally-confining $\U(N)$ vacua,

\begin{equation}
\label{ferrari:moduli}
u_p = \frac{N}{p} \sum_{q=0}^{\lfloor p/2 \rfloor} \comb{p}{2q} \comb{2q}{q} \Lambda^{2q} \phi^{p-2q} 
\end{equation}

If we wish, we can rewrite this expression in terms of the gaugino bilinear (``glueball") S, by performing a Legendre transformation with respect to $\Log(\Lambda^{2N})$ (\ie``integrating in S") \cite{Ferrari:2002jp}:

\begin{eqnarray}
\label{eq:glueballw}
W(\phi, g_p, S, \Lambda^2) &=& \sum_{p \ge 1} g_p u_p(\phi,\Lambda^2 = y) + S \Log(\frac{\Lambda^{2N}}{y^N}) \nonumber \\
&=& N \sum_{p \ge 1} \frac{g_p}{p} \sum_{q=0}^{\lfloor p/2 \rfloor} \comb{p}{2q} \comb{2q}{q} y^q \phi^{p-2q} + S \Log(\frac{\Lambda^{2N}}{y^N})
\end{eqnarray}
Of course, this form of $W$ does not contain any additional information, but it is useful for comparison with other techniques such as the matrix model (where $S$ is the fundamental variable of the matrix model perturbation expansion), and the analysis using integrable systems of \cite{Boels:2003fh}.  Indeed, we will recover this expression from the Lax matrix of the affine Toda system in section \ref{sec:intsys}.

\subsection{Factorization of the Seiberg-Witten curve for $N_f>0$}
\label{sec:nffactor}

The Seiberg-Witten curve for \ntwo\ gauge theory with $0 \le N_f  < 2 N_c$ fundamental hypermultiplets is \cite{D'Hoker:1997nv}

\begin{equation}
\label{eq:curvehypers}
y^2 = P_{N_c}(x)^2 - 4 \Lambda^{2N_c-N_f} \prod_{i=1}^{N_f} (x+m_i)
\end{equation}
where $m_i$ are the bare hypermultiplet masses.  When $N_f \ge N_c$ there is an ambiguity in the curve, and a polynomial of order $N_f - N_c$ in $x$ (multiplied by appropriate powers of $\Lambda$ to have well-defined scaling dimension $n$) may be added to $P_{N_c}(x)$ without changing the \ntwo\ prepotential.  We will mainly be interested in the case $N_f = N_c$, for which $P_{N_c}(x)$ is therefore ambiguous at the constant order proportional to $\Lambda^N$.

The curve (\ref{eq:curvehypers}) can be scaled to recover the $N_f=0$ curve (\ref{curvepure}) by taking the limit

\begin{equation}
\Lambda \rightarrow 0,\hspace{0.3in} m_i \rightarrow \infty,\hspace{0.3in}  \Lambda^{2N_c-N_f} \prod m_i \equiv \tilde{\Lambda}^{2N_c}
\end{equation}
with $\tilde{\Lambda}$ finite.  Note that the latter identification is the scale-matching relation of the theories above and below the mass scale of the fundamentals.

We now show how the factorization using Chebyshev polynomials can be generalized to the hypermultiplet curve (\ref{eq:curvehypers}) (this problem has been studied indirectly using matrix models in \cite{Demasure:2002jb}).  Define the functions

\begin{eqnarray}
P_{N_c}(\theta) &=& \sum_{i=0}^{N_f}  \nu_i \Cos((N_c-i) \theta) \nonumber \\
Q_{N_c}(\theta) &=& \imath \sum_{i=0}^{N_f} \nu_i \Sin((N_c-i) \theta)
\end{eqnarray}
Then
\begin{eqnarray}
P_{N_c}^2 - Q_{N_c}^2 &=& \sum_i \nu_i^2 + 2 \sum_{i \neq j} \nu_i \nu_j ( \Cos(i \theta) \Cos(j \theta) + \Sin(i \theta) \Sin(j \theta) )\nonumber \\
&=& \sum_i \nu_i^2 + 2 \sum_{i \neq j} \nu_i \nu_j \Cos((i-j) \theta) \equiv R_{N_f}(\theta)
\end{eqnarray}
Therefore the equation
\begin{equation}
P_{N_c}^2 - R_{N_f} = Q_{N_c}^2
\end{equation}
gives the desired factorization of the Seiberg-Witten curve by setting $\Cos(\theta)= \frac{x}{2 \tilde \Lambda}$ for $\U(N)$, or  $\Cos(\theta)= \frac{x- \Lambda}{2 \tilde \Lambda}$ for $\SU(N)$, where the shift is needed to cancel the $x^{N-1}$ term in $P_N(x)$.  The parameters $\nu_i$ are related to the fundamental masses $m_i$, although the relations are polynomial in general.

This expression simplifies dramatically when $N_f = N_c, m_i \equiv m$, and we find

\begin{eqnarray}
P_N &=& \sum_{i=0}^N \comb{N}{i}\beta^{N-i} \Cos(i \theta) \nonumber \\
&=& (\beta + e^{\imath \theta})^N + (\beta + e^{-\imath \theta})^N \nonumber \\
Q_N &=& \imath \sum_{i=0}^N \comb{N}{i}\beta^{N-i} \Sin(i \theta) \nonumber \\
&=& (\beta + e^{\imath \theta})^N - (\beta + e^{-\imath \theta})^N
\end{eqnarray}
with $\beta = \Lambda/\tilde{\Lambda}$, where $\Lambda$ is the scale of the theory with flavours, and $\tilde{\Lambda}^2 = m \Lambda$ is the parameter defined above that corresponds to the dynamical scale of the theory in the limit where the fundamentals have been scaled out completely.  If we choose a limit where the fundamental masses become very large compared to the scale $\Lambda$, \ie~$\beta$ becomes a small parameter, then the curve can be treated as a small deformation of the $N_f=0$ curve. 

After some algebra, we obtain the following expression for $P_N(x)$:

\begin{equation}
P_N(x) = 2 \Lambda^N \delta_{N,0} + \sum_{i=1}^N \frac{i}{2} \comb{N}{i} \Lambda^{N-i} \tilde \Lambda^i \sum_{r=0}^{\lfloor \frac{i}{2} \rfloor} \frac{(-1)^r}{i-r} \comb{i-r}{r} \frac{(x - \Delta)^{i-2r}}{\tilde \Lambda}
\end{equation}
where $\Delta = 0$ for $\U(N)$ and $\Delta=\Lambda$ for $\SU(N)$ to cancel the first subleading power of $x$.  We can resum this expression to extract the $s_k$.  For $\U(N)$ we find

\begin{equation}
\label{eq:sk}
s_{N-j} = \Lambda^{N-j} \sum_{r=0}^{\lfloor \frac{i-j}{2} \rfloor} (j+2r)\comb{N}{j+2r} \frac{(-1)^r}{j+r} \comb{j+r}{r} \left(\frac{\tilde \Lambda}{\Lambda}\right)^{2r} + 2 \Lambda^N \delta_{j,0}
\end{equation}
and for $\SU(N)$ we find

\begin{equation}
s_{N-j} = \Lambda^{N-j} \sum_{i=1}^N i \comb{N}{i}\sum_{r=0}^{\lfloor \frac{i-j}{2} \rfloor} \frac{(-1)^{i-j-r}}{i-r} \comb{i-r}{r} \comb{i-2r}{j}  \left(\frac{\tilde \Lambda}{\Lambda}\right)^{2r} + 2 \Lambda^N
\delta_{j,0}
\end{equation}

We now compare to the results obtained in section \ref{sec:uv} based on period integrals of the \nOne\ curve.  Recall that we obtained the expression (\ref{eq:generalnf}):

\begin{equation}
W = N_c b_0 + N_c \mu c_1 \Log(\frac{2 \tilde{\Lambda}}{\mu} ) + \sum_{i=1}^{N_f} [W(\xi(m_i))]_- 
\end{equation}
It can be verified in examples that for $N_f=N_c$, taking $\mu=2 \tLambda$ (so that it cancels the $\Log$ term) will extremize this superpotential.  We assume that this holds true for general $W(\Phi)$; this technique also works to extremize the $N_f=0$ superpotentials with arbitrary $W(\Phi)$, but does not seem to work for general $N_f$.  We obtain

\begin{equation}
\label{eq:uk}
W _{\hbox{low}}= N_c \left( \sum_{p=1}^{\lfloor \frac{n}{2} \rfloor} \frac{g_{2p}}{2p} \comb{2p}{p} \tilde \Lambda^{2p} + \sum_{i=1}^{n+1} \frac{g_i}{i} \Lambda^i \sum_{j=0}^{\lfloor \frac{i-1}{2} \rfloor} \comb{i}{j} \left( \frac{\tilde \Lambda}{\Lambda}\right)^{2r} \right)
\end{equation}
From this expression can be read off the values of the gauge-invariant moduli $\langle u_k \rangle = \frac{\partial W}{\partial g_k}$.  Note that our result has the form of a finite series expansion in $\beta$; in the limit $\beta=0$ we recover the superpotential of the $N_f=0$ theory $W_{\hbox{low}} = N b_0$.  The $u_k$ are related to the curve parameters $s_k$ via the Newton formula

\begin{equation}
k u_k + k s_k + \sum_{i=1}^{k-1} i u_i s_{k-i} = 0
\end{equation}

As in section \ref{sec:sw}, the $\SU(N)$ moduli $\tilde u_k$ may be obtained from the $\U(N)$ by shifting away the trace:

\begin{equation}
\label{eq:shiftedmoduli}
\tilde u_k = \sum_{i=1}^{N}(x_i - \frac{u_1}{N})^k
\end{equation}
Expanding the powers in (\ref{eq:shiftedmoduli}) one finds

\begin{equation}
\tilde u_k = \frac{1}{k} \left( \sum_{j=1}^k (\frac{-u_1}{N})^{k-j} j \comb{k}{j} u_j + N (\frac{-u_1}{N})^k\right)
\end{equation}

We have verified in a number of cases that the $u_k$ associated to the $s_k$ (\ref{eq:sk}) obtained from the factorized Seiberg-Witten curve agree with the values calculated from the superpotential (\ref{eq:uk}), up to a physically irrelevant sign $\Lambda \rightarrow - \Lambda$ (which can be absorbed into the conventions used to define the Seiberg-Witten curve (\ref{eq:curvehypers})) and the ambiguity in the top modulus $u_N$ at order $\Lambda^N$.

For example, the factorization for the first few $\U(N)$ curves is achieved by:
\\\\
$\U(2)$:
\begin{eqnarray}
P_2(x) = x^2 + 2x \Lambda - 2 \tLambda^2 + 2 \Lambda^2  \nonumber \\
u_1 = -2 \Lambda,\ u_2 = 2 \tLambda^2
\end{eqnarray}
$\U(3)$:
\begin{eqnarray}
P_3(x) = x^3 + 3 x^2 \Lambda + x(-3 \tLambda^2 + 3 \Lambda^2) - 6 \tLambda^2 \Lambda + 2 \Lambda^3 \nonumber \\
u_1 = -3 \Lambda,\ u_2 = 3(\tLambda^2 + \frac{1}{2} \Lambda^2),\ u_3 = 3(-\tLambda^2 \Lambda - \frac{2}{3} \Lambda^3)
\end{eqnarray}
$\U(4)$:
\begin{eqnarray}
P_4(x) = x^4 + 4 x^3 \Lambda + x^2(-4 \tLambda^2 + 6 \Lambda^2) + x(-12 \tLambda^2 \Lambda + 4 \Lambda^3) + 2 \tLambda^4 - 12 \tLambda^2 \Lambda^2 + 2 \Lambda^4 \nonumber \\
u_1 = -4 \Lambda,\ u_2 = 4(\tLambda^2 + \frac{1}{2}\Lambda^2),\ u_3 = 4(-\tLambda^2 \Lambda^2 - \frac{1}{3} \Lambda^3),\ u_4 = 4(\frac{3}{2} \tLambda^4 + \tLambda^2 \Lambda^2)
\end{eqnarray}
which can be compared to the $u_k$ read off from (\ref{eq:uk}):

\begin{equation}
u_1 = N \Lambda,\ u_2 = N (\tLambda^2 + \frac{1}{2} \Lambda^2),\ u_3 = N(\tLambda^2 \Lambda + \frac{1}{3} \Lambda^3),\ u_4 = N(\frac{3}{2} \tLambda^4 + \tLambda^2 \Lambda^2 + \frac{1}{4} \Lambda^4)
\end{equation}

\section{Effective superpotentials from integrable systems}
\label{sec:intsys}

We begin this section by proving that the superpotential calculation of \cite{Boels:2003fh} using the integrable structure of the \ntwo\ gauge theory (see \cite{D'Hoker:1999ft} for a  review), yields the same result in the maximally-confining phase as (\ref{eq:glueballlow}), (\ref{eq:glueballw})  obtained from the $N_f=0$ period integral and factorization calculations.

The integrable system associated to pure \ntwo\ Yang-Mills theory is the periodic Toda chain, with Lax matrix:
\begin{equation}
\label{eq:laxmatrix}
L = \left( \begin{array}{ccccccc}
\phi_1 & y_1 & 0 & \ldots &0 & z \\
1 & \phi_2 & y_2 & 0 & \ldots & 0 \\
0 & \ddots & \ddots & \ddots & \ddots & 0 \\
0 & \ddots & \ddots & \ddots & \ddots & 0 \\
0 & \ddots & \ddots & \ddots & \ddots & y_{N-1} \\
y_N/z & 0 & \ldots & 0 & 1 & \phi_N
\end{array}\right)
\end{equation}
The conserved quantities (Hamiltonians) of the Toda system $U_k = {\frac{1}{k+1}} \Tr L^k$ are associated to the gauge-invariant polynomials $u_k = {\frac{1}{k+1}} \Tr \Phi^k$ that parametrize the moduli space of the \ntwo\ gauge theory.  The spectral curve of the Lax system is defined by
\begin{equation}
{\hbox{det}}(x.I - L) \equiv P_N(x) + (-1)^N (z + \Lambda^{2N} z^{-1}) = 0
\end{equation}
Under the change of coordinates
\begin{equation}
\label{yz}
y = 2z + (-1)^N P_N(x)
\end{equation}
the spectral curve becomes
\begin{equation}
\label{spectral}
y^2 = P_N(x)^2 - 4 \Lambda^{2N}
\end{equation}
which is the other standard form of the Seiberg-Witten curve of \ntwo\ $\U(N)$ Yang-Mills theory.

Therefore, when we deform the \ntwo\ theory by turning on a tree-level superpotential
\begin{equation}
W = \sum_{i=1}^{n+1} g_i u_i
\end{equation}
the analogous quantity in the Toda system is the corresponding function of the conserved quantities $U_i$.  The essence of the proposal of \cite{Boels:2003fh} is that evaluating $W(L)$ gives the exact effective superpotential of the theory\footnote{When the superpotential $\Wtree$ contains terms of degree $N$ or higher, the spectral parameter $z$ that appears in the Lax matrix (\ref{eq:laxmatrix}) will appear in the $U_k$.  However, in the quantum \nOne\ gauge theory these moduli are ambiguous because the operators $\Tr \Phi^k, k \ge N$ receive quantum corrections, and the resolution proposed in \cite{Boels:2003fh} was that all occurrences of $z$ in the Lax superpotential $W(L)$ should be discarded at the end of the computation (alternatively they can be supressed to arbitrarily high orders by embedding $\U(N) \subset \U(tN)$).}.

We will now obtain the explicit form of $\WLax$ for a given $\Wtree$ and recover the result in section \ref{sec:sw}.  For this purpose the form of the Lax matrix (\ref{eq:laxmatrix}) is slightly awkward to work with, because the $z$ entries are not on the same footing as the other variables.  To rectify this, conjugate $L$ by ${\hbox{diag}}(1, z^{1/N}, z^{2/N}, \ldots, z^{N-1/N})$ to bring it into the form:

\begin{eqnarray}
\label{eq:todalaxy}
L &\sim& \left( \begin{array}{ccccccc}
\phi & \frac{y}{z^{1/N}} & 0 & \ldots &0 & z^{1/N} \\
z^{1/N} & \phi & \frac{y}{z^{1/N}} & 0 & \ldots & 0 \\
0 & \ddots & \ddots & \ddots & \ddots & 0 \\
0 & \ddots & \ddots & \ddots & \ddots & 0 \\
0 & \ddots & \ddots & \ddots & \ddots & \frac{y}{z^{1/N}} \\
\frac{y}{z^{1/N}} & 0 & \ldots & 0 & z^{1/N} & \phi
\end{array}\right) \nonumber \\
&=& \phi I + \frac{y}{z^{1/N}} S + z^{1/N} S^{-1}
\end{eqnarray}
where $S$ is the $N\times N$ shift matrix, satisfying $S^N = I$.

Therefore,
\begin{eqnarray}
\Tr(L^p) &=& \Tr\Big(\sum_{l=0}^p \phi^{p-l} \comb{p}{l} I \sum_{m=0}^l \left(\frac{y}{z^{1/N}}\right)^m z^{-m/N} S^{2m-l} \comb{l}{m} \Big) \nonumber \\
&=& N \sum_{l=0}^{p} \phi^{p-l} \comb{p}{l} {\sum_{a=-\lfloor \frac{l}{2N} \rfloor}^{\lfloor \frac{l}{2N} \rfloor}} y^{(Na+l)/2} z^{-a} \comb{2l}{\frac{Na+l}{2}}
\end{eqnarray}
where in the second line we have used the fact that the terms can only appear on the diagonal if $2m-l = Na$, $a \in \Z$.  Suppressing powers of $z$ whenever they appear, we obtain

\begin{equation}
\WLax= N \sum_{p \ge 1} \frac{g_p}{p} \sum_{q=0}^{\lfloor p/2 \rfloor} \comb{p}{2q} \comb{2q}{q} \phi^{p-2q} y^q + S \Log(\frac{\Lambda^{2N}}{y^N})
\end{equation}
which recovers the expressions (\ref{eq:glueballlow}), (\ref{eq:glueballw}) obtained using exact field theory techniques, and by evaluating period integrals.

\subsection{Towards a new Lax matrix for $N_f=N_c$}
\label{sec:laxnf}

The connection between \ntwo\ gauge theories and integrable systems can be summarized by identifying the matrix $\Phi$ of the quantum gauge theory with a Lax matrix for the integrable system.  Therefore, if we can evaluate $\Phi$ in a given vacuum, we know the value of the Lax matrix in an equilibrium configuration of the integrable system.  Knowing the values of the moduli $\langle u_k \rangle$ in the particular \ntwo\ vacuum gives $N_c$ equations for the matrix $\Phi$, which is enough in principle to determine $\Phi$ up to gauge transformations.

In the previous section, we showed how evaluating the Toda Lax matrix in a particular equilibrium configuration (all position and momentum variables equal, \ie~$\phi_i\ \equiv\ \phi,\\ y_i\ \equiv\ y\ \equiv\ \tLambda^2$) allows us to recover the $\langle u_k \rangle$ of the factorized Seiberg-Witten curve.  Conversely, given the $\langle u_k \rangle$, we can reconstruct the Lax matrix of the periodic Toda chain: the $\langle u_k \rangle$ may be obtained from the single matrix\footnote{The entry with coefficient 2 exists because (\ref{eq:todalax}) does not contain the spectral parameter $z$ (which does not have a physical meaning in the gauge theory), so we can absorb the entry $\tLambda^2/z$ of (\ref{eq:laxmatrix}) into this entry.}

\begin{equation}
\label{eq:todalax}
\langle \Phi \rangle = \left( \begin{array}{cccccc}
\phi & \tLambda^2 & 0 & \ldots & 0 \\
1 & \phi & \tLambda^2 & \ldots & 0 \\
\vdots & \ddots & \ddots &  \ddots & \vdots \\
0 & \ldots & 1 &\phi & 2 \tLambda^2 \\
0 & \ldots & 0 & 1 & \phi
\end{array}\right)
\end{equation}

One can explicitly see from this expression how the classical value of $\Phi = \hbox{diag}(\phi_1, \ldots, \phi_N)$ is deformed by quantum effects, specifically the interaction with the background magnetic field of the condensed monopoles, which generates the off-diagonal terms (this can most easily be derived via compactification to 3 dimensions, where the four-dimensional monopoles reduce to 3-dimensional instantons \cite{deBoer:1997kr}).

We follow the same philosophy for the $N_f=N_c$ vacua studied in section \ref{sec:nffactor}, and identify the matrix $\Phi$ from which the expectation values of the moduli $\langle u_k \rangle$ in the maximally-confining vacua may again be obtained by taking the trace of powers.  We therefore have a candidate for a Lax matrix of the associated integrable system, which in these examples are spin chains \cite{Gorsky:1996hs, Gorsky:1996qp}\footnote{A Lax matrix is known for these systems, but it is written in transfer matrix form, i.e~as a chain of coupled $2\times2$ matrices, so it is not clear how to use it in the manner of \cite{Boels:2003fh}.  However, since a given integrable system may have more than one Lax pair, and the matrices may even be of different rank, we should not be discouraged from looking for a new Lax formulation.}.

We find for $\SU(N_c)$, $N_f=N_c$ and all quark masses equal, that the moduli $\langle u_k \rangle$ of the maximally confining vacuum may be obtained from the matrix

\begin{equation}
\langle \Phi \rangle = \left( \begin{array}{ccccccc}
0 & \tLambda^2 & \Lambda \tLambda^2 &  \Lambda^2 \tLambda^2 & \ldots &  \Lambda^{N-2} \tLambda^2 &  N \Lambda^{N-1} \tLambda^2 \\
1 & 0 & \tLambda^2 & \Lambda \tLambda^2 & \ldots &\Lambda^{N-3} \tLambda^2  & (N-1) \Lambda^{N-2} \tLambda^2 \\
0 & 1 & 0 & \tLambda^2 &  \ldots & \Lambda^{N-4} \tLambda^2 &  (N-2) \Lambda^{N-3} \tLambda^2 \\
& \ddots & \ddots & \ddots & & \ddots & \vdots \\
0 & 0 & 0 & 1 & 0 & \tLambda^2 & 3 \Lambda  \tLambda^2 \\
0 & 0 & 0 & 0 & 1 & 0 & 2 \tLambda^2 \\
0 & 0 & 0 & 0 & 0 & 1 & 0 \\
\end{array} \right)
\end{equation}
Note that this reduces to the Toda Lax matrix (\ref{eq:todalax}) in the appropriate scaling limit $\Lambda \rightarrow 0$ ($\phi = 0$ for the $\SU(N)$ vacua to ensure tracelessness).

It remains to generalize this matrix to a general vacuum and to better understand the relationship with the degrees of freedom of the spin chain system.

\section{Conclusions}

We have shown that the technique of \cite{Cachazo:2001jy} for computing effective superpotentials in terms of period integrals of the reduced Seiberg-Witten (or matrix model) curve is regularized by embedding the gauge theory within the UV-finite theory with $2N_c$ fundamental quarks; this implements the period integral cut-off required to regularize the divergences of the integral.  We can obtain the effective superpotential for any value of $N_f < 2N_c$ by taking the quark masses to infinity and decoupling them from the low-energy theory.

The enumeration of planar diagrams with boundary (associated to gauge theory diagrams with quark propagators) reproduces the expression for the glueball superpotential in terms of period integrals on the \nOne\ curve, in agreement with existing geometrical techniques: the period integral (\ref{eq:quarkgenfn}) is the generating function of planar diagrams with one boundary.

We showed explicitly how to factorize the Seiberg-Witten curve for theories with fundamental hypermultiplets, in terms of Chebyshev polynomials; this generalizes the known $N_f=0$ result.  In certain cases the form of the factorized curve is particularly simple, and we obtain general formulae for the moduli of the factorized curves.

The underlying structure of the \ntwo\ quantum gauge theories we have studied (and the superpotentials of associated \nOne\ theories) is that the scalar component of the adjoint chiral superfield $\Phi$ parametrizes the vacua of the theory, and is identified with a Lax matrix of an associated integrable system.  We generalized the Lax matrix of the periodic Toda chain (associated to \ntwo\ pure gauge theory) to the theory with $N_f=N_c$ fundamental hypermultiplets of equal mass, in the maximally confining vacua.  It remains an open problem to generalize this new Lax matrix to arbitrary vacua of the $N_f=N_c$ theory.

\bigskip
{\bf Acknowledgements:} It is a pleasure to thank A.~Brandhuber, R.~Corrado, N.~Halmagyi, K.~Intriligator, C.~R\"omelsberger, C.~Vafa and V.~Yasnov for useful discussions.  K.K.~thanks the Perimeter Institute for Theoretical Physics for their hospitality while this work was being completed.
This work was supported in part by funds provided by the DOE under grant number DE-FG03-84ER-40168.

\bibliographystyle{utphys}
\bibliography{lax}

\end{document}